\documentclass[aps,prl,twocolumn,superscriptaddress,showpacs]{revtex4}

\usepackage{graphicx}
\usepackage{color}

\begin{document}
\title{Large tuneable Rashba spin splitting of a two-dimensional electron gas in Bi$_2$Se$_3$}

\author{P.~D.~C.~King}
\affiliation{School of Physics and Astronomy, University of St. Andrews, St. Andrews, Fife KY16 9SS, United Kingdom}

\author{R.~C.~Hatch}
\author{M.~Bianchi}
\affiliation{Department of Physics and Astronomy, Interdisciplinary Nanoscience Center, Aarhus University, 8000 Aarhus C, Denmark}

\author{R.~Ovsyannikov}
\affiliation{Helmholtz-Zentrum Berlin, Elektronenspeicherring BESSY II, D-12489 Berlin, Germany}

\author{C.~Lupulescu}
\affiliation{Helmholtz-Zentrum Berlin, Elektronenspeicherring BESSY II, D-12489 Berlin, Germany}
\affiliation{Technische Universit{\"a}t Berlin, Strasse des 17.\ Juni 135, D-10623 Berlin, Germany}

\author{G.~Landolt}
\author{B.~Slomski}
\author{J.~H.~Dil}
\affiliation{Physik-Institut, Winterthurerstr.\ 190, Universitat Z{\"u}rich-Irchel, 8057 Z{\"u}rich, Switzerland}
\affiliation{Swiss Light Source, Paul Scherrer Institut, CH-5232 Villigen, Switzerland}

\author{D.~Guan}
\affiliation{Department of Physics and Astronomy, Interdisciplinary Nanoscience Center, Aarhus University, 8000 Aarhus C, Denmark}
\affiliation{Department of Physics, Zhejiang University, Hangzhou 310027 China}

\author{J.~L.~Mi}
\affiliation{Center for Materials Crystallography, Department of Chemistry, Aarhus University, 8000 Aarhus C, Denmark}

\author{E.~D.~L.~Rienks}
\affiliation{Helmholtz-Zentrum Berlin, Elektronenspeicherring BESSY II, D-12489 Berlin, Germany}

\author{J.~Fink}
\affiliation{Helmholtz-Zentrum Berlin, Elektronenspeicherring BESSY II, D-12489 Berlin, Germany}
\affiliation{Leibniz-Institute for Solid State and Materials Research Dresden, P.O.Box 270116, D-01171 Dresden, Germany}

\author{A.~Lindblad}
\affiliation{MAX-lab, Lund University, P.O. Box 118, SE-22100 Lund, Sweden}

\author{S.~Svensson}
\affiliation{MAX-lab, Lund University, P.O. Box 118, SE-22100 Lund, Sweden}
\affiliation{Department of Physics and Astronomy, Uppsala University, P.O. Box 521, SE-75121 Uppsala, Sweden}

\author{S.~Bao}
\affiliation{Department of Physics, Zhejiang University, Hangzhou 310027 China}

\author{G.~Balakrishnan}
\affiliation{Department of Physics, University of Warwick, Coventry CV4 7AL, United Kingdom}

\author{B.~B.~Iversen}
\affiliation{Center for Materials Crystallography, Department of Chemistry, Aarhus University, 8000 Aarhus C, Denmark}

\author{J.~Osterwalder}
\affiliation{Physik-Institut, Winterthurerstr.\ 190, Universitat Z{\"u}rich-Irchel, 8057 Z{\"u}rich, Switzerland}

\author{W.~Eberhardt}
\affiliation{Helmholtz-Zentrum Berlin, Elektronenspeicherring BESSY II, D-12489 Berlin, Germany}
\affiliation{Technische Universit{\"a}t Berlin, Strasse des 17.\ Juni 135, D-10623 Berlin, Germany}

\author{F.~Baumberger}
\affiliation{School of Physics and Astronomy, University of St. Andrews, St. Andrews, Fife KY16 9SS, United Kingdom}

\author{Ph.~Hofmann}
\affiliation{Department of Physics and Astronomy, Interdisciplinary Nanoscience Center, Aarhus University, 8000 Aarhus C, Denmark}

\date{\today}
\begin{abstract}
We report a Rashba spin splitting of a two-dimensional electron gas in the topological insulator Bi$_2$Se$_3$ from angle-resolved photoemission spectroscopy. We further demonstrate its electrostatic control, and show that spin splittings can be achieved which are at least an order-of-magnitude larger than in other semiconductors. Together these results show promise for the miniaturization of spintronic devices to the nanoscale and their operation at room temperature.
\end{abstract}

\maketitle

Spintronics promises to revolutionize electronics and computing by making explicit use of the electron's spin in addition to its charge~\cite{Zutic:Rev.Mod.Phys.:76(2004)323--410}.
A key-requirement is spin manipulation via an electric, rather than a magnetic, field. However, even the prototypical method~\cite{Datta:Appl.Phys.Lett.:56(1990)665-667} for demonstrating such control, via an electrostatically-tuneable Rashba~\cite{Bychkov:JETPLett.:39(1984)78} spin splitting of a two-dimensional electron gas (2DEG), is difficult to implement: despite decades of research, crucial spintronic components such as the spin-field and spin-Hall effect transistors have only been realized very recently in a laboratory setting~\cite{Koo:Science:325(2009)1515--1518,Wunderlich:Science:330(2010)1801}, and are currently restricted to operation at cryogenic temperatures. A breakthrough is hampered by the intrinsic properties of available materials. In particular, the modest spin-orbit interaction of most semiconductors leads to small Rashba splittings, necessitating low temperatures for device operation and long channel lengths with ultra-high purity material to avoid spin-flip scattering events.

States with larger Rashba splittings are known to exist at several metal surfaces~\cite{LaShell:Phys.Rev.Lett.:77(1996)3419--3422,Koroteev:Phys.Rev.Lett.:93(2004)046403,Ast:Phys.Rev.Lett.:98(2007)186807} and in ultrathin metal films~\cite{Frantzeskakis,He}. 
Unlike a semiconductor 2DEG, however, their spin splitting cannot be influenced by external electric fields, rendering such systems unsuitable for real devices. Here, we obtain a Rashba splitting of a semiconductor 2DEG at the surface of the topological insulator Bi$_2$Se$_3$. Through electrostatic control, we achieve spin-splittings comparable to those of metal surface states and surface alloys, and orders of magnitude larger than in other semiconductors. This system should therefore provide an ideal platform to establish spintronic control in practical devices. 

{\it In-situ} cleaved single-crystal samples of Bi$_2$Se$_3$ and Ca-doped Bi$_2$Se$_3$ were measured using conventional, time-of-flight, and spin-resolved angle-resolved photoemission (ARPES). The time-of-flight measurements were performed using a VG Scienta ArTOF 10k spectrometer~\cite{Oehrwall:J.ElectronSpectrosc.Relat.Phenom.:xx(2010)xxx-xxx} at BESSY II with the storage ring operating in single bunch mode at a repetition rate of 1.25~MHz. Conventional ARPES measurements were performed using a SPECS Phoibos 150 hemispherical analyser at the ASTRID synchrotron~\cite{Hoffmann:NuclearInstrumentsandMethodsinPhysicsResearchSectionA_523:441--453(2004)}. The spin-resolved measurements were performed using the COPHEE setup at the Swiss Light Source~\cite{Hoesch}. The energy and angular resolution was better than 5--15~meV (80~meV) and 0.13--0.2$^\circ$ (1.5$^\circ$) for the spin-integrated (resolved) measurements, respectively.

\begin{figure}
\begin{center}
\includegraphics[width=\columnwidth]{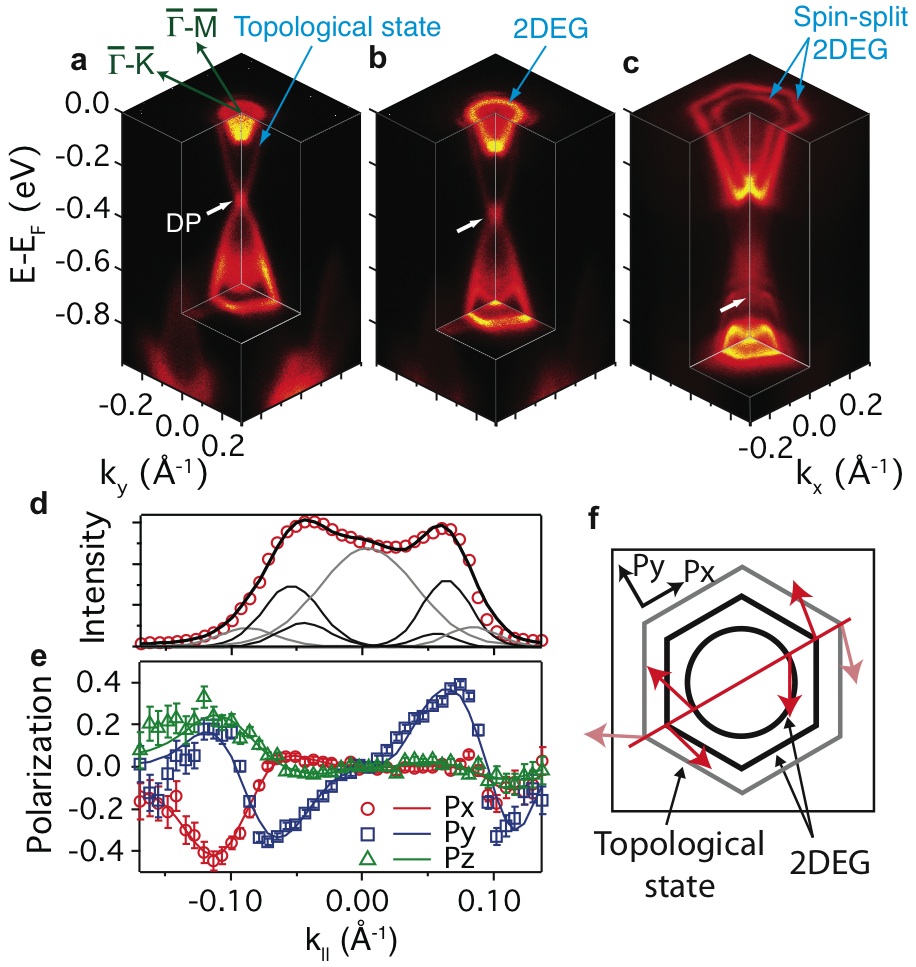}
\caption{ \label{f:time_dep} (a--c) Electronic structure of Bi$_2$Se$_3$, measured using time-of-flight ARPES at increasing times (marked in Fig.~\ref{f:time_dep_B}(a)) after cleaving the sample ($h\nu=19.2$~eV, $T=10$~K). (d) Spin-ARPES total intensity MDC at the Fermi level ($h\nu=19.5$~eV, $T=60$~K) along the $\bar{\Gamma}$--${\bar{M}}$ direction, and (e) the measured (points) and fitted (lines) radial ($P_x$), tangential ($P_y$), and out-of-plane ($P_z$) components of the spin polarization along this cut. (f) Schematic representation of the in-plane component of the Fermi surface spin texture.}
\end{center}
\end{figure}
The freshly-cleaved sample (Fig.~\ref{f:time_dep}(a)) shows the familiar linearly-dispersing topological surface state, whose apex defines the Dirac point, as well as the bulk valence bands at higher binding energy. Close to the Fermi level, occupied bulk conduction band states are also visible, due to the usual degenerate electron doping of this compound. With time, the Dirac point shifts to higher binding energies (Fig.~\ref{f:time_dep}(b)), indicating a downwards shift of the electronic bands close to the surface, associated with an increased surface electron doping~\cite{Bianchi:NatureCommun.:1(2010)128,Hsieh:Nature:460(2009)1101--1105}. The resulting quantum well causes a 2DEG to form at the surface~\cite{Bianchi:NatureCommun.:1(2010)128}, with two-dimensional subbands giving rise to the well-defined Fermi surface in Fig.~\ref{f:time_dep}(b).

With further surface doping (see Supplementary Movie 1~\cite{EPAPS}), the band bottom of the 2DEG state starts to shift away from the $\bar{\Gamma}$-point. This is indicative of a Rashba spin splitting: in the presence of a potential gradient (provided here by the band bending), spin-orbit coupling lifts the spin-degeneracy of the 2DEG, separating the states into two bands. In the simplest approximation, their dispersion is given by
\begin{equation}E^\pm(k_{\parallel})=E_0+\frac{\hbar^2k_{\parallel}^2}{2m^*}\pm\alpha{}k_{\parallel},\end{equation}
where $m^*$ is the effective mass and $\alpha$ is the Rashba coupling parameter, dependent on both the gradient of the potential and the spin-orbit coupling strength. The spin-split bands cross at $k_{\parallel}=0$, causing the band bottom to shift away from the zone centre as observed here. As the band bending, and consequently the near-surface potential gradient, continues to increase, a well defined spin splitting of the states is established (Fig.~\ref{f:time_dep}c), much like for the model example of the Au$(111)$ surface state~\cite{LaShell:Phys.Rev.Lett.:77(1996)3419--3422}. We stress, however, that the states we observe here are not surface states~\cite{Wray:NaturePhys.:7(2011)32--37}, but rather quantum-confined conduction band states, that is, a 2DEG similar to those found at a number of semiconductor surfaces and interfaces~\cite{King:Phys.Rev.Lett.:101(2008)116808,Ando:Rev.Mod.Phys._54:437--672(1982)}. However, the size of the splitting is much larger. Even for semiconductors generally considered to have a strong spin-orbit coupling, such as InAs and InSb, no Rashba splitting of a 2DEG could be resolved by previous ARPES studies~\cite{King:Phys.Rev.Lett.:104(2010)256803,Aristov:Appl.Surf.Sci._166:263--267(2000)}. Indeed, the momentum (energy) splitting at the Fermi level of up to  $\Delta{k}_F=0.08$~\AA$^{-1}$ ($\Delta{E}_R=180$~meV) are approximately one to two orders of magnitude  larger than in semiconductor 2DEGs which have previously been considered good candidates for spintronic devices~\cite{Engels:Phys.Rev.B:55(1997)R1958--R1961,Nitta:Phys.Rev.Lett.:78(1997)1335--1338,Grundler:Phys.Rev.Lett._84:6074--6077(2000)}, and a factor of 2-4 larger than for the Au$(111)$ surface state~\cite{LaShell:Phys.Rev.Lett.:77(1996)3419--3422,Nechaev:Phys.Rev.B:80(2009)113402}.
 
Apart from expected matrix-element effects~\cite{Bianchi:NatureCommun.:1(2010)128}, neither the topological state nor the spin-split states vary when measured with different photon energies (Supplementary Movie~2~\cite{EPAPS}). This lack of $k_z$ dispersion confirms the two-dimensional nature of the electron gas that gives rise to the Rashba splitting. Further, we have confirmed their spin-split nature by spin-resolved ARPES. The spin polarization of the bands crossing the Fermi level along the ${\bar\Gamma}$--${\bar{M}}$ direction (Fig.~\ref{f:time_dep}(d)) were extracted (Fig.~\ref{f:time_dep}(e)) using the procedure described in Ref.~\cite{Meier2008}. The total intensity momentum distribution curve (MDC) was fit with 7 components, accounting (from high $|k_{||}|$) for the topological state (gray lines) and the spin-split 2DEG state (black lines), with a single central component to account for the residual bulk conduction band and unresolved shallow 2DEG states. The analysis shows that both the topological and the spin-split 2DEG states are 100\% spin polarized and the spin texture of the latter is that expected for a simple Rashba model. The spins lie predominantly in-plane, and averaging over the Kramers pair are situated at an angle of $\sim\!-95^\circ$ ($\sim\!86^\circ$) relative to the state's momentum. The deviations from perfect spin-helicity most likely arise from a slight sample misalignment, although we cannot rule out spin-canting due to hexagonal warping of the Fermi surface. However, a detailed analysis of such features is beyond our current experiment. The spin of the topological state is aligned approximately anti-parallel to that of the outer Rashba-split branch of the 2DEG state, and possesses finite out-of-plane spin polarization, possibly~\cite{Gedik} due to hexagonal warping of this state once the 2DEG starts to develop~\cite{Bianchi:NatureCommun.:1(2010)128}. The in-plane spin texture around the Fermi surface extracted from these measurements is summarized schematically in Fig.~\ref{f:time_dep}(f).

\begin{figure}
\begin{center}
\includegraphics[width=\columnwidth]{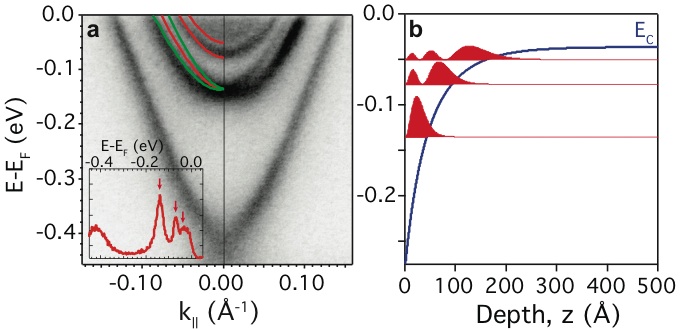}
\caption{ \label{f:model} (a) Detailed measurement ($h\nu=16$~eV, $T=60$~K) of multiple 2DEG states (upper features) and the topological surface state (lower feature). Model Poisson-Schr{\"o}dinger calculations (red solid lines) reproduce the ladder of 2DEG states within the surface quantum well. Inset: EDC at $\bar{\Gamma}$ with 2DEG states marked by arrows. (b) Near-surface band bending (blue) with the calculated energy levels and modulus-squared wave functions of the quantum-confined states (red). A Rashba splitting of $\alpha=0.36$~eV{\AA} (estimated from the $k_F$-splitting) applied to the lowest calculated subband dispersion gives the green lines in (a).}
\end{center}
\end{figure}
Tuning matrix elements via the photon energy, we can resolve additional shallow states which are also non-dispersive in $k_z$ (Supplementary Movie~2~\cite{EPAPS}). From a detailed measurement (Fig.~\ref{f:model}(a)), a clear second state can be discerned above the spin-split band bottom, while a very weak and diffuse third state can just be distinguished close to the bulk conduction band minimum. In agreement with earlier studies~\cite{King:Phys.Rev.Lett.:104(2010)256803,Bianchi:NatureCommun.:1(2010)128,Meevasana:NatureMater.:10(2011)114--118}, we attribute these to a ladder of quantum-well subband states induced by the surface band bending. To support this assignment, we have performed Poisson-Schr{\"o}dinger calculations, following the method described in Refs.~\cite{King:Phys.Rev.B:77(2008)125305,Bianchi:NatureCommun.:1(2010)128}. Taking the total band bending magnitude as the shift of the Dirac point from its position in the freshly cleaved sample, and taking the same materials parameters detailed in Ref.~\cite{Bianchi:NatureCommun.:1(2010)128}, we find quite good agreement with the experimental data. This agreement is further improved by a small reduction of the static dielectric constant to 70, yielding the subband structure shown in Fig.~\ref{f:model}, strongly supporting the multiple quantum-well-state origin of the observed bands. The change in dielectric constant compared to our earlier work~\cite{Bianchi:NatureCommun.:1(2010)128} may result due to the increased free-carrier density within the 2DEG established here, a factor which is known to cause a large reduction of dielectric constant in other semiconductors~\cite{eps_n}.

Only the lowest of these measured subbands exhibits a visible spin splitting in the ARPES. In a simple model developed to describe Rashba splitting of III-V semiconductor quantum wells~\cite{deAndradaeSilva:Phys.Rev.B:50(1994)8523--8533}, the Rashba coupling parameter
\begin{equation}\label{e:R1}\alpha\propto\left\langle\psi(z)\left|\frac{d}{dz}\left[\frac{1}{\varepsilon'(z)}-\frac{1}{\varepsilon'(z)+\Delta}\right]\right|\psi(z)\right\rangle,\end{equation}  
where $\varepsilon'(z)=\varepsilon+V(z)+E_g$, $\varepsilon$ is the subband energy relative to the bulk conduction band, $V(z)$ is the band bending potential, $E_g$ the band gap, and $\Delta$ is a measure of the spin-orbit coupling within a Kane $\mathbf{k}\cdot\mathbf{p}$ approach~\cite{Kane:J.Phys.Chem.Solids._1:249(1957)}. Consequently, for the deepest subband, whose wave function is localized close to the surface in the region of steepest potential (Fig.~\ref{f:model}(b)), a much stronger Rashba splitting would be expected than for the shallower states, whose wave functions are much more extended in regions of shallower potential. Indeed, taking a representative value of $\Delta=1$~eV, this model predicts Rashba parameters of 0.1, 0.035, and 0.014~eV\AA{} for the three subbands. This underestimates the observed splitting of the lower subband by a factor of $\sim\!4$, which is not unexpected for a simple model intended to describe standard semiconductor materials with quite different bulk band structure and much weaker spin-orbit interactions. However, the qualitative agreement with the trend of spin splittings between the different subbands strongly supports that the Rashba splitting here is dominated by the gradient of the back-side of the potential well.   

\begin{figure}
\begin{center}
\includegraphics[width=\columnwidth]{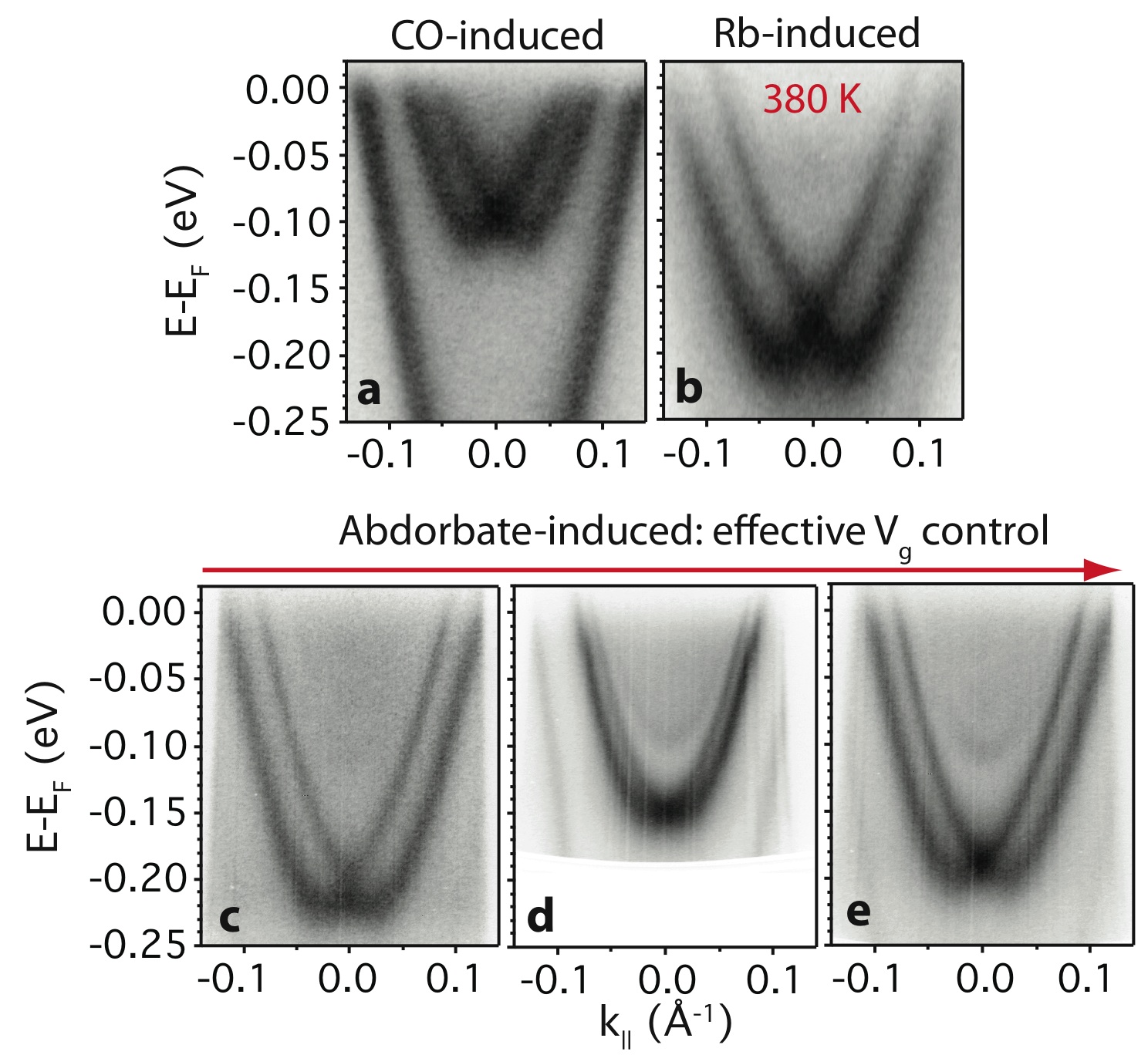}
\caption{ \label{f:control} Rashba-split 2DEG in Ca-doped (non-degenerate) Bi$_2$Se$_3$ induced by (a) $18$~L of CO-adsorption at $60$~K and (b) $\ll1$~ML of Rb-deposition at 170~K. The measurement temperature in (a) and (b) is 60~K and 380~K, that is, well above room temperature, respectively.(c--e) Reversible nature of effective gating of undoped Bi$_2$Se$_3$, with the spin splitting (along $\bar{\Gamma}$--$\bar{K}$) changing from (c) $\Delta{k_F}=0.034$~\AA$^{-1}$ ($V_g\sim\!0.28$~eV) to (d) $\Delta{k_F}=0.011$~\AA$^{-1}$ ($V_g\sim\!0.11$~eV) following annealing the sample at 250~K to remove some surface adsorbates to (e) $\Delta{k_F}=0.025$~\AA$^{-1}$ as the effective gate potential is increased again to $V_g\sim\!0.26$~eV.  }
\end{center}
\end{figure}

Most likely, the 2DEG formation here is mediated by the adsorption of residual gas molecules such as CO from the vacuum onto the surface. Indeed, deliberate dosing with CO leads to much more rapid formation of the 2DEG~\cite{Bianchi_CO} (Fig.~\ref{f:control}(a)), while its formation can also be accelerated via deliberate adsorption of H$_2$O~\cite{Benia_H2O}, another common residual gas in ultra-high vacuum. The spin-split 2DEG can also be induced by deposition of small quantities ($\ll1$~ML) of alkali metals on the surface (Fig.~\ref{f:control}(b)), as well as by deposition of other donor species such as Fe~\cite{Wray:NaturePhys.:7(2011)32--37} or Cu~\cite{Wray_Cu}.
Irrespective of the exact microscopic origin of the surface donors, their presence is analogous to application of an external gate field. Indeed, via removal of surface adsorbates by mild annealing (250~K), we reduce this effective gate field, showing that the spin splitting can correspondingly be diminished. This demonstrates the reversible and controllable nature of the Rashba splitting, driven by a change in the band bending (Fig.~\ref{f:control}(c--e)). We note that this is fundamentally different from the adsorbate-induced changes to Rashba splittings achieved in conventional surface states, as here it reflects the electrostatically-tuneable nature of a semiconductor 2DEG. In a practical device, of course, this would be achieved by electrically-biasing a gate electrode. The injected charge necessary to induce the strong Rashba splittings observed here may be achievable using conventional dielectric media, where substantial gate-tuned changes in Hall concentration have already been demonstrated for Bi$_2$Se$_3$ thin films~\cite{Chen:Phys.Rev.Lett.:105(2010)176602}. Alternatively, the sheet densities required are well within the range that have recently been realized in electric-double-layer FETs~\cite{Ueno:NatureMater.:7(2008)855--858}, which is directly analogous to the surface alkali-metal doping shown in Fig.~\ref{f:control}(b).

\begin{figure}
\begin{center}
\includegraphics[width=\columnwidth]{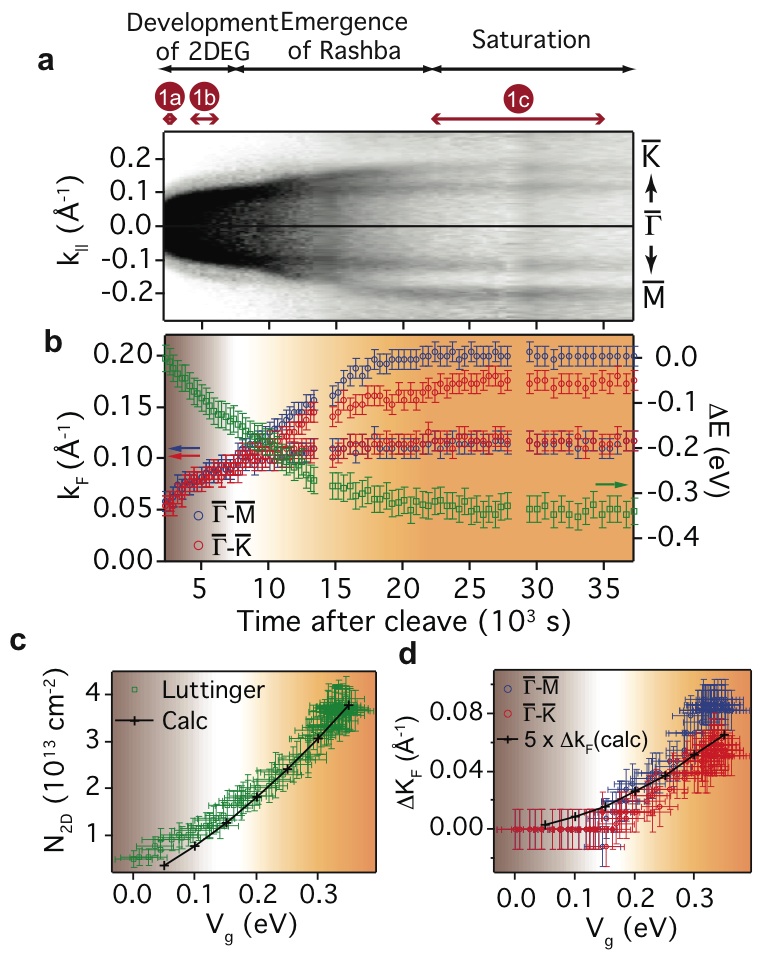}
\caption{ \label{f:time_dep_B} (a) Evolution of the Fermi wave vector along $\bar{\Gamma}-\bar{M}$ and $\bar{\Gamma}-\bar{K}$ with time after cleaving the sample. (b) Extracted $k_F$ positions and band-bending which induces the 2DEG. (c) Sheet density and (d) $k_{\parallel}$-splitting at the Fermi level of the measured (points) and calculated (lines) Rashba-split 2DEG as a function of the effective gate potential, $V_g$.}
\end{center}
\end{figure}
Using surface adsorbates to achieve this field-effect control uniquely permits the simultaneous spectroscopic investigation of the 2DEG. Utilizing time-of-flight ARPES~\cite{Oehrwall:J.ElectronSpectrosc.Relat.Phenom.:xx(2010)xxx-xxx}, we track the continuous evolution of the full $(k_x,k_y,E)$-dependent electronic structure as a function of effective gate voltage (Fig.~\ref{f:time_dep_B} and Supplementary Movie~1~\cite{EPAPS}).
With time, the effective gate potential, given by the shift of the Dirac point, monotonically increases in magnitude up to $\sim\!0.35$~eV. Concurrent with this, the Fermi surface of the 2DEG grows and smoothly evolves into two spin-split sheets above an effective gate voltage of $\sim\!0.15$~eV. Crucially, through a combination of different sample temperatures, background pressure, and differing times after the cleave, the effective gate potential reaches a factor of three higher than in our previous study~\cite{Bianchi:NatureCommun.:1(2010)128}, a necessary condition for realization of the Rashba split states we report here. From these measurements, the sheet density of the 2DEG (Fig.~\ref{f:time_dep_B}(c), obtained from its Luttinger area) and the momentum splitting at the Fermi level (Fig.~\ref{f:time_dep_B}(d)) can be extracted as a function of the effective gate voltage. The measured sheet density and trend in the spin splittings are well reproduced by our model calculations, again supporting the importance of the potential gradient of the backside of the quantum well in determining the Rashba splitting.

By tuning this gradient by an effective gate voltage change of only $\sim\!200$~mV, the spin splitting can be varied from zero up to $\sim\!0.08$~\AA$^{-1}$ (Fig.~\ref{f:time_dep_B}(d), corresponding to a Rashba parameter of $\sim\!1.3$~eV{\AA}. This is orders of magnitude larger than the control of spin-splitting achieved in conventional semiconductor 2DEGs~\cite{Nitta:Phys.Rev.Lett.:78(1997)1335--1338}, making Bi$_2$Se$_3$-based 2DEGs ideal for applications where electrical manipulation of the spin precession is required. For example, the precessional phase shift for injected electrons in the spin-FET is $\Delta{k}_FL$, where $L$ is the channel length~\cite{Datta:Appl.Phys.Lett.:56(1990)665-667}. Consequently, for the 2DEG created here, the required channel length for maximal source-drain current modulation could be as small as $\sim4$~nm, providing a route towards nanoscale spintronic devices. Furthermore, the large Rashba effects ensure that a well defined spin-splitting is maintained up to well above room temperature, as shown in Fig.~\ref{f:control}(b). This is a necessary requirement for practical implementation of spintronics, but something which has yet to be demonstrated for devices based on more conventional semiconductors~\cite{Koo:Science:325(2009)1515--1518,Wunderlich:Science:330(2010)1801}. The recent achievements of electrostatic gating~\cite{Chen:Phys.Rev.Lett.:105(2010)176602,Steinberg:NanoLett.:10(2010)5032--5036} of Bi$_2$Se$_3$ as well as its  epitaxial growth on Si~\cite{Zhang:Appl.Phys.Lett.:95(2009)053114} and GaAs~\cite{Richardella:Appl.Phys.Lett.:97(2010)262104} indicate good prospects to incorporate the spintronic functionality suggested here into practical devices, and to integrate these with conventional semiconductor electronics.

\

Support from the UK EPSRC, the Scottish Funding Council, the ERC, the Danish Council for Independent Research (Natural Sciences), the Lundbeck foundation, the Swedish Research Council (VR), Vinnova in Sweden, the Carl Tryggers foundation for Scientific Research, PSI (SLS), the Swiss National Science Foundation, and the EU is gratefully acknowledged. M.~Lundqvist and P.~Karlsson (VG Scienta AB) are thanked for assistance with the new ArTOF 10k spectrometer.

\end{document}